\documentclass{aastex}

\shorttitle{Nuclei Ring of NGC 4314}
\shortauthors{Ann}

\begin{document}

\title{HYDRODYNAMIC SIMULATIONS FOR THE NUCLEAR MORPHOLOGY OF NGC~4314}

\author{ H. B. Ann}
\affil{Department of Earth Sciences, Pusan National University, Pusan
   609-735, Korea}
    
\email{hbann@cosmos.es.pusan.ac.kr}

\begin{abstract}
We performed SPH simulations to study the nuclear morphology of a barred 
galaxy NGC~4314. We have constructed the mass models based on the 
results of a profile decomposition into disk, bulge, and bar components.
Our models have three different nuclear structures according to the 
assumption about the nuclear bar: no nuclear bar, a synchronous nuclear bar
and a fast nuclear bar. Our SPH  simulations show that 
the morphology of the nuclear region of NGC~4314
which is characterized by an elongated ring/spiral of newly formed stars
and HII regions, aligned nearly parallel to the primary bar can be understood
in terms of the secular evolution driven by the non-axisymmetric potential.
The slightly elongated and aligned nuclear ring of NGC~4314 can be formed
by the strong barred potential and the moderate central concentration of
the bulge mass with and without a nuclear bar. However, the nuclear 
spiral pattern can not be developed without a nuclear bar. 
The nuclear bar of NGC~4314 seems to rotate faster than the primary bar
since the nuclear morphology induced by the synchronous nuclear bar is much
different from the observed one. 
\end{abstract}

\keywords{galaxies: morphology, SPH simulations, NGC 4314} 

\section{INTRODUCTION}

The nuclear ring of a barred spiral galaxy NGC4314 has been studied 
extensively because of 
its peculiar morphology in the nuclear region as clearly shown in the Hubble
atlas \citep{san61}.  Earlier photometric studies showed that the color
of the nuclear ring is bluer than that of the surrounding regions including
the nucleus \citep{lyn73,ben80}.  The blue color
of the nuclear ring has been ascribed to the young stellar populations that
might have formed from the disk material driven into a nuclear region 
by the bar  potential \citep{gar91,ben92,ben93,ben96}.
The presence of the young stellar populations in the nuclear
region of NGC~4314 was suggested by the emission lines in the central
region \citep{bur62,wak80} and the nuclear
spirals in the the center of the bar \citep{lyn74,ben80}. Recent
$HST$ images of NGC~4314 clearly show that the nuclear ring is a nuclear
spiral which consists of recently formed star clusters
and $HII$ regions \citep{ben93}. 

Thanks to extensive numerical studies, we are beginning to understand the
formation and evolution of nuclear rings in barred galaxies.
The presence of the ILRs is required for 
the formation of a nuclear ring \citep{sch81,sch84,com85}.
The gravitational torque of bar on the
gas induces gas inflow which leads to the formation of a nuclear ring
at ILRs \citep{com96}. Because the shape and orientation of a nuclear 
ring depend on the dynamical properties of a 
galaxy \citep{com85,ath92,pin95,ann00},
we can use the morphology
of a nuclear ring, combined with hydrodynamic simulations, as a probe of 
the underlying dynamics. 
A general analysis of gas responses to bar forcing in disk galaxies
is given by \citet{ann00}.

For NGC~4314 seems to be a classic example of a barred galaxy whose
bar drives disk gas into a ring at ILRs, we aim to analyze the morphology
and dynamics of the central region of NGC~4314 by SPH simulations.
In \S2, we describe the global and nuclear morphology of NGC~4314.   
Models of the present simulations are presented in \S 3 and the results of 
SPH simulations are described in \S 4.
Discussion on the nuclear ring morphology and nuclear bar is given in \S 5
and the conclusions are given in the last section.

\section{MORPHOLOGY}
\subsection{Global Morphology}

NGC~4314 is an early type barred galaxy, classified as SBa.  
The global morphology of NGC~4314 is characterized by a prominent bar
and a peculiar bulge with faint outer spiral arms.
There are many photographs and CCD images in the literature which show
the general morphology of NGC~4314 but the global morphology is best seen 
in the Hubble atlas \citep{san61}.
NGC~4314 is a nearly face-on galaxy with inclination of about $23^{\circ}$.
The bulge is somewhat elongated along the bar and it seems
to be triaxial because it shows isophotal twists.  The length, axial ratio,
and position angle of the bar are $\sim 66^{\prime \prime}$, $\sim 4$,
and $\sim 148^{\circ}$, respectively \citep{ben96,ann99}. 
Because the position angle of the major axis of the disk is about
$59^{\circ}$ \citep{ben96}, the bar aligns almost perpendicular to the disk.
The outer spiral arms
are very faint but they trace $130^{\circ}$ arcs out to $125^{\prime \prime}$
from the nucleus \citep{gar91}.  
There seems to be no HII
region in the outer arms \citep{bur62}.  The faintness of 
the outer spiral arms is due to lack of gas \citep{gal75,gar91}
which prevents recent star formation in the outer disk.  However, the main 
three components, that is, 
disk, bulge, and bar have comparable luminosities \citep{ann99}.

\subsection{Nuclear Morphology}

NGC~4314 has been known to have a peculiar morphology in the central
part of the galaxy \citep{san61}.  Earlier
studies of NGC~4314 \citep{ben80,gar91,ben92,ben93} 
showed a nuclear ring and nuclear spiral pattern 
along with several hot spots in the nuclear region.  
Fig.~1 shows a deprojected grey scale image of the nuclear region of 
NGC~4314 captured by $HST/NICMOS$ with F160W filter \citep{mul01}. 
The integration time of
the image is 360s. The deprojection of the image was made by 
using the disk inclination of $23^{\circ}$ \citep{ben96}.
We adjust the contrast of the image to get the most clear picture of the
nuclear features. 
We see clearly that the nuclear ring of NGC~4314 is a nuclear spiral which
consists of young stellar clusters of newly formed massive stars and dust
lanes. The dust lanes of spiral pattern extend further than the spiral arms
and obscure some parts of the nuclear ring.
The diameter of the nuclear ring is $\sim 16^{\prime \prime}$. It is
elongated close to the bar axis which is indicated by the dotted line 
in Fig.~1. Note, however, that they are not perfectly aligned,
$\Delta \theta \approx 10^{\circ}$.  
We plot the major axis of the disk as a solid line for comparison.

Along with the nuclear ring, there seems to be an oval or
a nuclear bar at the center of the galaxy \citep{ben92,ben93}.
The first estimate of the diameter of the nuclear bar is about 
$3^{\prime \prime}$ \citep{ben92}, but 
\citet{ben96} adopted the diameter of the nuclear bar
as $8^{\prime \prime}$ from the $HST$ image of NGC~4314 \citep{ben93}.
To analyze the geometrical properties of the nuclear features, we applied
an ellipse fitting to the isophotes of the deprojected image of NGC~4314.
Fig.~2 shows the profiles of the luminosity, ellipticity, and position 
angle of the nuclear region of NGC~4314. 
The luminosity profile which is dominated by the $r^{1 \over 4}$-law bulge 
is very smooth except for the regions affected by the dust lanes and
the nuclear ring. The  nuclear bar contributes an appreciable amount of
luminosity which makes a shallow gradient from $r \approx 4^{\prime \prime}$
to the end of the nuclear bar.
As indicated in Fig.~2, the nuclear bar ends at $r \approx 5^{\prime \prime}.2$
and the locations of the dust lanes and
the nuclear ring are $\sim 7^{\prime \prime}$ and $\sim 8^{\prime \prime}$,
respectively. Our estimate of the nuclear bar length agrees with
that of \citet{ben96} if we take into account the projection effect.

As shown in Fig.~2, the nuclear bar and the nuclear ring show excess of
luminosity above the bulge luminosity while the dust lanes show 
deficiency of luminosity.
The gradual increase of the ellipticity within $r \approx 5.^{\prime \prime}5$
is due to the nuclear bar. The complicated variation of the ellipticity
profile reflects the complex structure of the nuclear region.
The ellipticity of the  nuclear ring is about 0.35,
which is slightly larger than that of the dust lanes. The position angles
of the nuclear features are very similar with 
$\Delta \theta \lesssim 5^{\circ}$. 
 
Although the presence of a nuclear bar inside the nuclear ring of
NGC~4314 has been suggested \citep{ben92, ben93, ben96},
its morphology is not known yet due to the
dominant bulge luminosity.  To see the nuclear bar more clearly,
we subtracted the axisymmetric bulge component from the deprojected image
of NGC~4314 by assuming that the bulge luminosity follows
the de Vaucouleurs's $r^{1 \over 4}$ law. To do this,
we derived the effective brightness and effective radius of NGC~4314
from the minor axis profile which is little affected by the luminosity 
of the nuclear bar.  Fig.~3 shows the bulge subtracted image of the nuclear
region of NGC~4314 where we see clearly the nuclear bar that aligns almost
parallel to the nuclear ring. We plot an ellipse whose ellipticity
and position angle are determined from the ellipse fitting employed in Fig.~2.
We see that the ellipse fits 
nicely the nuclear ring. The brightest hot spots outside the ellipse
belong to the nuclear spiral arms. 

\section {MODELS}

The basic numerical method employed in the present study is the 
Smoothed Particle Hydrodynamics (SPH) technique which is known to 
be a powerful 
tool for a wide variety of astrophysical problems (see \citep{mon92} for
a thorough review).  We used the same code as that of \citet{ann00}.
We simulate the responses of gaseous 
disk by SPH particles distributed uniformly in the beginning of each
simulation, with circular velocities for the centrifugal equilibrium
to the gravitational accelerations.  
The initial radius of the gaseous disk is 4.5 kpc.
The number of SPH particles is around 10,000, which gives initial
resolution length of about 130 pc.
Throughout the calculations, we fixed
the temperature of the gaseous disk at 10,000 K, because radiative
cooling is effective in the shocked gas.
For simplicity and better resolution, our calculations were confined to
two-dimensional disk.

\subsection {The Potential}

We assumed that the galactic potential, which is due to three stellar
components (bulge, disk, and bar) and the dark halo, is independent of time
in the frame corotating with the bar.
We include the dark halo to match the observed flat rotation 
curve in the outer part \citep{com92}. We have performed model simulations
with and without a nuclear bar to see the effect of the  nuclear bar 
on the response of gaseous disk. 
We used the same functional forms of the potentials of disk, bulge,
bar, and dark halo as those of \cite{lee99}. In brief, we assumed the
exponential disk \citep{fre70}, Plummer's model, bi-axial potential of
\citet{lon92}, and logarithmic potential for disk, bulge, bar, and
dark halo, respectively.
We add the bar potential slowly over half a bar
revolution ($\tau_{bar} \approx 10^{8} yr$) to avoid the violent shocks due to
a sudden non-axisymmetric force.
We ignore the self-gravity of the 
gaseous disk because its mass is assumed to be a small fraction of the
total mass of the model galaxy.
The effect of self-gravity is most pronounced
in the central region where the density of SPH particles is high due to
the bar-driven inflow \citep{ann00}.

\subsection{Model Parameters}

The free parameters that define the gravitational potential of each
component can be constrained by the mass distributions inferred from
the luminosity distributions of NGC~4314.  We derived the mass fractions
and scale lengths of the disk, the bulge, and the bar from the results
of a profile decomposition of the $V$-band surface photometry \citep{ann99}.
Here we assumed a constant mass-to-luminosity ratio throughout the galaxy.
Only exception to this is the bulge scale length because there is no 
direct match between the scale parameters of the adopted functional
form for the bulge potential and those of the assumed function
of the bulge luminosity distribution. 
We also assumed a large extended halo whose
mass inside the bar radius is about 10 \% of the visible mass. 
The length of the nuclear bar is assumed to be $\sim 10^{\prime \prime}$
with its axial ratio of 4. 
We have assumed the mass of the nuclear bar as 2 \% of the primary bar
based on its size. 

The pattern speed of the bar $\Omega_{p}$ is constrained by the resonance
locations and the maximum rotational velocity that is determined by the
radius and mass scale of the galaxy. 
The length scale of our models is taken as the radius of bar, $R_{sc}$=3 kpc,
assuming the distance of NGC~4314 as 10 Mpc. 
We assumed that the total visible mass within
the bar radius is $2 \times 10^{10}M_{\odot}$ which is about 10 times larger
than the total mass interior to $r=450$ pc \citep{gar91}
and one fourth of the dynamical mass of NGC~4314 \citep{sag93}.
We assumed the mass of the gaseous disk as 1 \% of the total
visible mass, $M_{g}=2 \times 10^{8}M_\odot$, which is virtually the same 
as the mass of molecular gas ($2.1 \times 10^{8}M_\odot$) derived
by \citet{ben96}. With these radius and mass
scales, the dynamical time scale $\tau_{dyn}$ is $1.7\times10^7$ yr and
the unit of angular frequency $\Omega_{sc}$ is 56.6 km/sec/kpc. 
Assuming that the bar ends near corotation ($R_{CR} \approx 1.2 R_{bar}$),
we determined the pattern speed of the bar as $\Omega_{p}=33.6$ km/sec/kpc
which corresponds to 0.6 $\Omega_{sc}$. Our bar pattern speed is almost
the same as that derived by the resonance model of \citet{gar91}.
We list the adopted parameters of the models in Table~1. A-model does not
have a nuclear bar while B- and C-model have a synchronous nuclear bar and
a fast nuclear bar, respectively. 
The pattern speed of the fast nuclear bar
was adopted as $\Omega_{s} \approx 12\Omega_{p}$ to ensure that the 
CR of the nuclear bar coincides with the location of the IILR of the 
primary bar.  

Fig.~4 shows the rotation curve and angular frequencies of the simplest
mass model (A-model). 
The shape of the rotation curve is similar to those derived from the
$K$-band image \citep{com92} and the CO observations \citep{ben96}. 
However, our rotation curve rises somewhat more slowly than those
of \citet{com92} and \citet{ben96} but it rises more
steeply than that of \citet{qui94} which is obtained from a mass model
based on a $K$-band image.  The maximum rotational velocity of the model
galaxy is about 180 km/sec which is in a good agreement with \citet{com92}
and \citet{ben96}. We indicated the
resonance locations in Fig.~4, where IILR, OILR, and OLR stand for the inner
inner Lindblad resonance, the outer inner Lindblad resonance, and the
outer Lindblad resonance, respectively.  The pattern speed of the bar
is represented as the dot-dashed line.
Because the mass of the nuclear bar is assumed to be negligible,
the rotation curves of the models with nuclear bars (B- and C-model) are
virtually the same as that of the simplest model.
The rotation curves of our mass models result in a general agreement
between the locations of the IILR of the model galaxies and the observed
radius of the nuclear ring of NGC~4314. 

\section {Results}
\subsection {Evolution of Global Morphology}
Since the global morphological evolutions of the gaseous disk of the models
with and without nuclear bars are very similar, we present the evolution
of the simplest model which has no nuclear bar (A-model) in Fig.~5 to see 
the general gas responses 
which lead to the development of a nuclear ring, dust lanes and faint outer 
spiral arms. The number in the upper left corner of each panel represents
the evolution time in unit of bar revolution time $\tau_{bar}$ that 
corresponds to $\sim 6.4 \tau_{dyn}$.
In the early phase of evolution, a two-armed spiral pattern of density
enhancement is developed in the gaseous disk due to bar forcing that
drives gas streaming motion. The gas inside the CR moves inward due to
the loss of angular momentum to the bar, while the gas outside the CR moves
outward by gaining angular momentum from the bar.  The spiral pattern inside
the CR transforms into a perpendicular nuclear ring near OILR with highly
curved symmetric dust lanes within one bar revolution. 
While the misalignment of the nuclear ring is decreasing
and the dust lanes become less curved, the outer spiral arms extend further.
The dust lanes eventually turn into the centered straight dust lanes by the
time when the nuclear ring aligns to the bar after $\sim 2\tau_{bar}$.

As the evolution proceeds, the dust lanes become weaker and the size of the 
nuclear ring decreases with increasing density until they reach a steady 
state after $\sim 8\tau_{bar}$.  The location and morphology of the nuclear 
ring after $\sim 5\tau_{bar}$ resembles those of the nuclear ring of 
NGC~4314. The faint outer arms are reminiscent pseudo outer 
rings $R^{\prime}$ observed in barred galaxies \citep{but95}. Most of the 
gas within the CR is accumulated to the nuclear ring at the IILR but some 
fraction of gas come across the nuclear ring and 
falls into the nucleus itself.  

Fig.~6 shows a snap shot of the velocity field of the gaseous disk of A-model
at $t=4 \tau_{bar}$. The arrows represent the directions of the
particle velocities with the lengths being proportional to the velocities.
As is evident in the figure, the SPH particles near the CR have almost
zero velocity.
The streaming motion of gas particles is easily seen inside the
CR where gas particles lead the bar. The streaming motion indicates that 
there is a large amount of gas inflow along the bar. The velocity of
inflowing gas is up to $\sim 100$ km/sec which agrees with the observations
of \citet{ben96}.
The outer trailing spiral arms are composed of gas particles
that move outward by gaining angular momentum
from the bar while the straight dust lanes and ring-like structures
are composed of gas particles that move inward by losing their angular momentum 
to the bar.  It is also worth noting that the slightly curved leading dust
lanes, seen in Fig.~5, are the regions of density enhancement with 
abrupt velocity changes.
The high density and the sudden velocity changes may imply that
they are the loci of shock fronts. In later times, when the evolution of the 
gaseous disk reaches steady state, the streaming motion which leads
to the gas inflow toward the nucleus is significantly reduced.

\subsection{Evolution of Nuclear Region}

Although the global morphological evolutions of the gaseous disks are
not much different among models with or without nuclear bars,
the evolutions of the nuclear features are quite different due to
the perturbations by nuclear bars. 
Fig.~7 shows the evolution of the central part of the gaseous disk of the 
models with a synchronous nuclear bar (B-model) and a fast nuclear bar with
$\Omega_{s} \approx 12 \Omega_{p}$ (C-model),
along with that of the simplest model.
The CR of the nuclear bar is located near the IILR of the 
primary bar for the C-model.
Because the gas responses outside the nuclear ring 
are almost identical in every evolution stage,
we plot the nuclear regions only for better resolution. 
The box size of Fig.~7 is 1.2 kpc in one dimension.
For easy comparisons of the models with observations, we plot 
the distribution of the SPH particles with the ellipse representing the
nuclear ring of NGC~4314.  
Because the primary bar lies along the horizontal axis in Fig.~7,
we rotate the ellipse accordingly.

The morphology of the inner parts
is similar to each other until $\sim 2\tau_{bar}$ but
they evolve differently afterwards. 
The nuclear ring-like structure developed in A model evolves to
a nuclear ring after $\sim 5\tau_{bar}$  whose general morphology 
resembles the nuclear ring morphology of NGC~4314. The nuclear ring consists
of several fragments of various morphology.
However, the spiral pattern observed in the nuclear ring of NGC~4314
can not be reproduced by A-model. The evolution of B-model shows a highly 
elongated nuclear spiral at $t \approx 2\tau_{bar}$ with a large 
accumulation of gas in the nucleus.
The nuclear spiral pattern evolves to a nearly circular nuclear ring located
far inside the IILR. There is not much gas in the nuclear ring owing to 
the large amount of infall to the nucleus. The steady state morphology
of the  nuclear ring of B-model is much different from that of NGC~4314.
The nuclear ring-like structure of C-model evolves to several fragments 
which distribute randomly inside the IILR. But after $\sim 5\tau_{bar}$,
they evolve to a two-armed spiral pattern with a large accumulation of gas
near the nucleus.  The morphology of the nuclear spiral at $\sim 8\tau_{bar}$
resembles the nuclear ring morphology of NGC~4314. However, the nuclear 
spiral developed in C-model is a transient one. Moreover, NGC~4314 has
no nuclear structure, similar to that developed in the nucleus of C-model.

There are some similarities in the evolution of the nuclear morphologies
among models. All the models show shrinking of nuclear rings with 
increasing of gas density inside due to the gas inflow toward the nucleus.
But, as evident from Fig.~7, while most of the gas inside the CR 
is accreted between the two ILRs in A-model, a large fraction of
the gas inside the CR moves into the nucleus after $\sim 2\tau_{bar}$ in 
B-model and after $\sim 3\tau_{bar}$ in C-model, respectively. 
One interesting point to be noted is the development of several fragments
which are thought to be the regions of density enhancement. Since there is
no common property in the  structures of the fragments, they might be
caused by some sort of dynamical instability due to random shocks.

\section {DISCUSSION}

\subsection{Orientation of Nuclear Ring}

Most of the nuclear rings observed in real galaxies are misaligned, usually
perpendicular to the primary bars \citep{but93}. Thus, the nuclear ring
of NGC~4314 which is elongated to the direction nearly parallel to the bar axis
with the ellipticity of $\epsilon \sim 0.35$ 
is a rare example of aligned nuclear rings. The misalignment angle
of the nuclear ring with the primary bar is only $\sim 10^{\circ}$.
It is well known that the morphology of nuclear rings depends on the 
dynamical properties of the host galaxies such as 
mass distribution and bar pattern speed.
The dependence of the nuclear ring morphology on the bar pattern speed 
was examined by \citet{com85} who showed that slow bars induce 
elongated nuclear rings aligned parallel to the bar while fast bars
that allow only one ILR generate circular nuclear rings. 

However, there seems to be not much freedom for the pattern speed of a bar
because it ends near the CR \citep{con80,ath92}. More important 
parameter that constrains the morphology of nuclear rings is the degree of
central mass concentration \citep{pin95,ann00}. 
\citet{pin95} showed that galaxies with high central
concentration and small bar axial ratio have circular nuclear rings,
while galaxies with low central concentration and large axial ratios
have elongated nuclear rings that tend to be aligned with the bars.
The nuclear rings of galaxies which have strong bars with axial ratio of 
${a \over b} \gtrsim 4$, however, can be aligned perpendicular to 
the bar if the central concentration is very high \citep{ann00}.
Thus, the central concentration of mass distribution seems to be the
primary parameter which controls the nuclear ring morphology.
The aligned nuclear ring of NGC~4314 is due to the strong bar with
a moderate degree of the central concentration of mass.

The central concentration is not an observable
quantity but it can be inferred from the rising part of the rotation curve
which is mainly determined by the mass distribution of the bulge component.
Centrally concentrated bulges give rise to steep rotation curves which
allow two ILRs unless the bar pattern speed is unrealistically high.
This is the reason for the preponderance of occurrence of nuclear rings
in early type galaxies \citep{but93}. 
In the present mass models, the rising part of the rotation curve strongly
depends on the bulge scale length $r_{b}$ in the Plummer spherical
potential. We adopted the bulge scale $r_{b}= 510$ pc, which is slightly
larger than that used by \citet{gar91} who modeled
NGC~4314 by a Plummer spherical bulge and Toomre disk with
bulge scale length of 400 pc.

\subsection {Shrinking of Nuclear Ring}

One interesting phenomenon observed in the evolution of the nuclear 
morphology is the shrinking of ring-like structures which are formed near
OILR. Their sizes are
reduced as much as the ratio of the $R_{OILR}$ to $R_{IILR}$.
Such a shrinking is not a common property of nuclear ring evolution.
It requires an extremely strong non-axisymmetric potential by  which 
the particles populating the aligned nuclear ring lose angular momentum
due to the gravitational torque. Misaligned nuclear rings or aligned 
nuclear rings with weak bars do not show such a shrinking \citep{ann00}. 
The strong non-axisymmetric potential of NGC 4314 is provided by the
large axial ratio (${a \over b} \approx 4$) and the high mass 
fraction ($\sim 30$\%) of the primary bar.

Shrinking of the nuclear ring
of NGC~4314 is suggested by \citet{com92} from the location of the 
molecular ring observed by a high resolution CO(J=1-0) mapping of NGC 4314.
They interpreted the mismatch of the locations of the radio ring and 
the molecular ring as a result of the shrinking of the molecular ring
which is located inside the radio ring. The position of the radio ring
coincides with the optical nuclear ring. 
However, the time-scale for the shrinking of the molecular ring is 
very short, comparable to the star formation time-scale of a few $10^{7}$ yr
\citep{com92}. In the models with nuclear bars the ring-like structures
formed at the OILR shrink to the IILR in a few $10^{8}$ yr, while in the 
model with no nuclear bar the shrinking time scale is about  $10^{9}$ yr.

The slow shrinking of the nuclear ring,  
as compared with the star formation time-scale,
indicates that the mean age of the stellar populations 
in the nuclear region changes with radius, 
older age near the OILR and younger age near the IILR,
if the nuclear ring of NGC 4314 has evolved from the ring-like structure
formed near the OILR. Thus, the variation of stellar colors,
progressively becoming redder with increasing distance from the nuclear ring,
observed by \citet{ben92} can be explained by the age difference of the 
stellar populations, as suggested by \citet{ben96}.

\subsection {Pattern Speed of Nuclear Bar}

As shown in Fig.~3, there seems to be no question about the presence of 
a nuclear bar inside the nuclear ring of NGC~4314. Its morphology is
similar to that of the primary bar. However, there is no information
about the dynamical properties of the nuclear bar except for the
small misalignment between the two bars, $\Delta \theta \approx 5^{\circ}$.
In a double barred system, from a dynamical point of view,
a favorable situation consists of a faster nuclear bar with its CR at the
location of the ILR of the primary bar when the primary bar ends near the
CR of the primary bar \citep{pfe90,fri93}. But, in the case of NGC~4314,
there is a possibility of the synchronous rotation
due to the virtual alignment of the two bars \citep{ben96}.

However, as evident from Fig.~7, the evolution of gas response 
inside the OILR depends sensitively on the pattern speed of the nuclear bars.
The model with a fast nuclear bar drives a complicated evolution which
leads to the formation of a two-armed nuclear spiral resembling the
nuclear ring of NGC~4314, while the model with a synchronous nuclear bar
develops a smaller and rounder nuclear ring than that of NGC~4314.
The spiral pattern developed in the early phase of evolution in the model
with a synchronous nuclear bar is very short-lived and highly elongated.
Thus, the nuclear bar of NGC~4314 is thought to rotate faster than the
primary one. We might observe the nuclear bar when it lies close to the
primary bar. However, since there is little gas in the nucleus of
NGC~4314 \citep{gar91}, there should be some mechanisms which remove
the infalled gas inside the nuclear ring if the nuclear structure of NGC~4314
can be explained by the models with a fast nuclear bar.

\section {CONCLUSIONS}

The peculiar nuclear morphology of an early type strong barred galaxy
NGC 4314 can be explained by dynamical models in which mass models are 
constrained by the results of profile decomposition and with the assumption
of $R_{CR} \approx 1.2 R_{bar}$ for the bar pattern speed.
Our SPH simulations well reproduce the shape and orientation of
the nuclear ring of NGC~4314 which is characterized by an
elongated ring and spiral arms of young stellar populations, 
partly obscured by the dust lanes and aligned nearly parallel to the
bar ($\Delta{\theta}\approx 10^{\circ}$). The aligned nuclear ring of
NGC 4314 can be formed from the gas inflow along the bar due to the moderate
central concentration of the bulge under the strong barred potential.
The constraints of the location of the nuclear ring 
at the IILR ($R_{IILR} \approx 350$ pc) and
the bar ends near the CR admit a slow rotation
of the bar ($\Omega_{p} \approx 34$ km/sec/kpc) which is effective for
the mass inflow toward the nucleus.

We have presented the results of three SPH simulations with different 
assumptions about
the nuclear structure; no nuclear bar, a synchronous nuclear bar, and
a fast nuclear bar. There is little difference in the evolutions of the 
global morphology among models but the evolutions of nuclear morphology 
are quite different due to the different nuclear dynamics. In the beginning
of the evolution of the gaseous disks, the nuclear ring forms from the density
enhancement of spiral pattern which is originated from the streaming 
motions due to the gravitational torque on the gas. At the time of
nuclear ring formation, it is aligned perpendicular to the primary bar
but the misalignment angle decreases after one bar revolution.
The diameter of the nuclear ring decreases as the inflowing gas accumulates
in the IILR with gradual disappearance of dust lanes along the bar.

If there is no nuclear bar, the nuclear spiral pattern similar to that of
NGC~4314 can not be developed even though the general morphology of the
nuclear ring located near the IILR resembles the nuclear ring of NGC~4314.
The synchronous nuclear bar drives a large infall into the nucleus and
develops a highly elongated nuclear spiral at earlier times which evolves
to a circular nuclear ring far inside the IILR after $\sim 5\tau_{bar}$. 
The steady state morphology induced by the synchronous nuclear bar is
much different from the observed one. 
In the case of the model with a fast nuclear bar ($\Omega_{s} \approx
12 \Omega_{p}$), the evolution of the nuclear feature is very complicated
due to the fluctuating potential by the fast nuclear bar. But, it leads to
the formation of a nuclear ring of spiral pattern which resembles
the observed nuclear ring of NGC~4314.
Thus, the nuclear bar of NGC~4314 seems
to rotate faster than the primary bar. The small misalignment between
the two bars, $\Delta \theta \approx 5^{\circ}$ might indicate that
we see the nuclear bar of NGC~4314 when it lies close to the primary bar.
However, there should be some mechanisms by which the infalled gas is 
removed from the nucleus if the nuclear morphology of NGC~4314 can be
explained by the nuclear dynamics induced by the fast nuclear bar. 

\acknowledgements
I thank Dr. Hyesung Kang for the help in the numerical works and 
Dr. Hyung Mok Lee for careful reading of the manuscript and discussion.
I would like to thank Dr. J. Mulchaey for providing me the HST/NICMOS
image of NGC 4314 before publication. I want to thank the anonymous referee
for the kind comments by which this paper is much improved.
This work was supported by the Interdisciplinary Research
Program of KOSEF through grant No. 1999-2-113-001-5.

\clearpage
\newpage
\begin{table}
\begin{center}
{\bf Table 1.}~~ Model parameters.
\vskip 0.3cm
\begin{tabular}{cccccccccccccccc}\hline
Model & $M_g$ & $r_h$ & $r_{d}$ & $r_{b}$ & a & b & $\Omega_p$& $M_d$& $M_b$& $M_{bar}$ & $M_s$ & $a_s$& $b_s$ & $\Omega_s$ \\ \hline
A  &   0.01& 3.0 & 1.0 & 0.17 & 1.0&  0.25&  0.6& .34&  .35&  .30&   \\
B  &   0.01& 3.0 & 1.0 & 0.17 & 1.0&  0.25&  0.6& .34&  .35&  .30& 0.005& 0.08& 0.02& 0.6& \\
C  &   0.01& 3.0 & 1.0 & 0.17 & 1.0&  0.25&  0.6& .34&  .35&  .30& 0.005& 0.08&
0.02& 7.0& \\
\hline
\end{tabular}
\end{center}
\tablenotetext{}{The parameters of the nuclear bars are designated by the 
subscript $s$.}
\end{table}

\newpage

\figcaption{annfig1.gif}{Deprojected $HST/NICMOS$ image of NGC 4314. We see
that the nuclear ring of NGC~4314 is a nuclear spiral which consists of
young stellar populations and dust lanes. The dust lanes of spiral pattern
extend further than the spiral arms and obscure some parts of the nuclear
ring. The nuclear ring is elongated close to the major axis of the primary
bar which is indicated as a dotted line. The  direction of the major axis 
of the disk is shown by a solid line. The oval shape inside the nuclear
ring is due to a nuclear bar which aligns almost parallel to the 
nuclear ring. The image size is $19.^{\prime \prime}2 \times 
19.^{\prime \prime}2$. North is up and East is to the left.
\label{fig1}}

\newpage

\figcaption{annfig2.gif}{Luminosity, ellipticity, and position angle profiles of 
the nuclear region of NGC 4314. The luminosity profile is dominated by the
bulge luminosity but the nuclear bar contributes an appreciable amount of
luminosity which makes the shallow gradient from $r \approx 4^{\prime \prime}$
to the end of the nuclear bar at $r \approx 5.^{\prime \prime}2$. 
We designate the locations of the end of the nuclear bar, the dust lanes,
and the nuclear ring. The gradual increase
of the ellipticity within $r \approx 5.^{\prime \prime}5$
is due to the nuclear bar. The position angles of the nuclear features
are similar but the small mismatch ($\Delta \theta \approx 5^{\circ}$)
between the orientations of the nuclear bar and the nuclear ring is 
worth to be noted.
\label{fig2}}

\newpage
\figcaption{annfig3.gif}{The deprojected and bulge subtracted image of the 
nuclear region of NGC~4314.  We see clearly a nuclear bar inside the
nuclear ring. We plot an ellipse which was derived from the
ellipse fit employed in Fig.~2. Most of the hot spots which are thought
to consist of the nuclear ring are located along the ellipse. We adjust
the display level to see the nuclear features most clearly.
The image size is $19.^{\prime \prime}2 \times
19.^{\prime \prime}2$. North is up and East is to the left.
\label{fig3}}

\newpage
\figcaption{annfig4.gif}{Rotational velocity and angular frequencies of the
mass models of NGC~4314. The mass fractions and the scale lengths are
constrained by the results of the profile decomposition \citep{ann99}.
The horizontal line represents the bar pattern speed $\Omega_{p}$ 
which is determined by the assumption of $R_{CR}\approx 1.2R_{bar}$.
We indicate the four resonance positions.
The unit of R is 3 kpc and the units of the velocity and the angular
frequency are 170 km/sec and 56.6 km/sec/kpc, respectively. 
\label{fig4}}

\newpage
\figcaption{annfig5.gif}{Evolution of the gaseous disk of the simplest 
model (A-model).
At earlier times the nuclear ring shows a large misalignment with the 
bar but becomes aligned parallel to the bar after $\sim 3 \tau_{bar}$.
The diameter of the nuclear ring decreases as it evolves.
The number of each panel is the evolution time in 
unit of $\tau_{bar}$ ($\sim 1.1 \times 10^{8}$ yr).
It reaches to the steady state after $\sim 8\tau_{bar}$.
The bar lies horizontally.  The box size is 12 kpc in one dimension.  
\label{fig5}}

\newpage
\figcaption{annfig6.gif}{Velocity field of A-model at $t=4\tau_{bar}$. 
The lengths of arrows are proportional to the particle velocities.  
The particles inside the corotation radius show strong streaming
motions along the bar. See the abrupt change of the particle velocity
near the bar axis, which might represent the loci of the shock fronts.
The length unit is 3 kpc for both axes.
\label{fig6}}

\newpage
\figcaption{annfig7.gif}{Evolution of the nuclear region of three models. 
A-model has no nuclear bar and B- and C-model have a  synchronous nuclear
bar and a fast nuclear bar ($\Omega_{s} \approx 12\Omega{p}$), respectively.
The ellipse which represents the nuclear ring of NGC~4314 is plotted
for easy comparison. Because the primary bar lies horizontally,
we rotate the ellipse accordingly. We see that there is no spiral pattern
in A-model, while B- and C-model show nuclear spiral pattern, at least
in some stages of evolution. Note that the nuclear bars induce much infall
into the nucleus.  The number of each panel represents the evolution time
in unit of $\tau_{bar}$. The box size is 1.2 kpc in one dimension.
\label{fig7}}

\end{document}